# Physical property and electronic structure characterization of bulk superconducting $Bi_3Ni$

Jagdish Kumar[1,2], Anuj Kumar[1], Arpita Vajpayee[1], Bhasker Gahtori[1], Devina Sharma[1,3], P.K. Ahluwalia[2], S. Auluck[1], and V.P.S. Awana[1,*]

[1]Quantum Phenomena and Application Division, National Physical Laboratory (*CSIR*)

New Delhi-110012, India

[2]Department of Physics, Himachal Pradesh University, Summerhill, Shimla-171005, India

[3]Department of Physics, Punjab University, Chandigrah 160014, India

We report the experimental and theoretical study on magnetic nature of $Bi_3Ni$ system. The structure is found to be orthorhombic (*Pnma*) with lattice parameters $a$ = 8.879Å $b$ = 4.0998Å and $c$ = 4.099Å. The title compound is synthesized via a solid state reaction route by quartz vacuum encapsulation of 5N purity stoichiometric ingredients of Ni and Bi. The superconducting transition temperature is found to be 4.1 K as confirmed from magnetization and specific heat measurements. The lower critical field ($H_{c1}$) and irreversibility field ($H_{irr}$) are around 150 and 3000Oe respectively at 2K. Upper critical field ($Hc_2$) as determined from in field (up to 4 Tesla) ac susceptibility is found to be around 2 Tesla at 2K. The normal state specific heat is fitted using Sommerfeld-Debye equation $C(T) = \gamma T + \beta T^3 + \delta T^5$ and the parameters obtained are $\gamma$= 11.08mJ/mol-K$^2$, $\beta$= 3.73mJ/mol-K$^4$ and $\delta$= 0.0140mJ/mol-K$^6$. The calculated electronic density of states (*DOS*) at Fermi level $N(E_F)$ and Debye temperature $\Theta_D$ are 4.697 states/eV per formula unit and 127.7K respectively. We also estimated the value of electron phonon coupling constant (λ) to be 1.23, which when substituted in MacMillan equation gives $T_c$ = 4.5K. Density functional (*DFT*) based calculations for experimentally determined lattice parameters show that Ni in this compound is non-magnetic and ferromagnetic interactions seem to play no role. The Stoner condition $I*N(E_F)$ = 0.136 per Ni atom also indicates that system cannot have any ferromagnetism. The fixed spin moment (*FSM*) calculations by fixing total magnetic moment on the unit cell also suggested that this system does not exhibit any signatures of ferromagnetism.

---

Key Words: $Bi_3Ni$, Structure, Magnetization, Superconductivity, Heat Capacity, and Density functional calculations.
PACS: 74.25.-q, 74.62.Bf, 74.62.Dh

[*]Corresponding Author
Dr. V.P.S. Awana
Fax No. 0091-11-45609310: Phone No. 0091-11-45608329
e-mail-awana@mail.nplindia.ernet.in; www.freewebs.com/vpsawana/

## Introduction

These are rich days for the search of new superconducting materials. This is because often new superconducting compounds keep on adding, which provide clues to the theoreticians to solve the puzzle of high $T_c$ superconductivity beyond 40K. The quest slightly speeded up once again (post high $T_c$ superconductivity (*HTSc*) avalanche [1]) with the discovery of superconductivity in $MgB_2$ [2] and Fe based oxy-pnictides [3]. The $T_c$ of latest entrants i.e., oxy-pnictides was raised by applying pressure ($T_c^{onset}$ = 43 K) [4] or substitution of smaller rare earth ion for the La site ($T_c$ = 55 K for $SmFeAsO_{1-x}F_x$) [5]. Soon after researchers reported other crystal structure types of layered Fe compounds to be superconducting, including $Ba_{1-x}K_xFe_2As_2$ [6], $Li_{1-x}FeAs$ [7] and FeSe [8]. Fe based oxy-pnictides crystallize in ZrCuSiAs-type structure with space group P4/nmm [3-5]. Very recently, a similar structure (ZrCuSiAs-type) Ni containing compound ($CeNi_{0.8}Bi_2$) is reported [9] from the consortium of the oxy-pnictide inventors [3] group. At this stage, we focused on another interesting Ni containing Bi compound i.e., $Bi_3Ni$, which is though known to be superconducting since 1951 [10], but only a few scant reports [11-13] do exist on its superconductivity. Tunneling spectroscopy results revealed that $Bi_3Ni$ is a strongly coupled superconductor [12]. The latest one exists in year 2000 by Y. Fujimori et al. [13]. More recently $Bi_3Ni$ is synthesized by vacuum encapsulation technique and possibility of co-existence of magnetism and superconductivity in this compound is addressed [14]. Keeping this in view, we synthesized $Bi_3Ni$ via an easy and versatile synthesis route, instead of conventional "arc melting". The studied compound is synthesized by sealing in quartz and vacuum annealing at 1000$^o$C for 24 hours and slowly cooled to room temperature. This is one shot heat treatment and post annealing etc. is not required. The resultant compound is superconducting below 4.1K and the Ni is non-magnetic. The non magnetic character of Ni in $Bi_3Ni$ is similar to that as observed earlier for the famous Ni Boro-Carbides viz. Y-Ni-B-C [15] and $MgCNi_3$ [16]. Detailed specific heat analysis and *DFT* calculations revealed that this compound is having electron-phonon coupling constant of around 1.23 and Ni is non magnetic. Further it is concluded that ferromagnetic interactions play no role in superconductivity of $Bi_3Ni$. This is in contrast to a recent report [14] related to possibility of coexistence of superconductivity and magnetism in $Bi_3Ni$. Our results will surely attract more researchers to work on superconductivity of this and similar Ni containing compounds.

## Experimental

Samples of nominal composition $Bi_3Ni$ are synthesized by solid state reaction route. The stoichiometric amounts of high purity (> 5N) Bi and Ni are ground thoroughly with help of an agate and mortar for around an hour. The mixed and pulverized powders are pressed into the form of a rectangular bar and are encapsulated in an evacuated quartz tube. The encapsulated tube was then heated at 1000$^o$C for over 24 hours and slowly cooled to room temperature. The X-ray diffraction patterns of the samples were obtained with the help of a Rigaku diffractometer using Cu$K_\alpha$ radiation. All physical property measurements including magneto-transport $R(T)H$, thermoelectric power $S(T)$, heat capacity $C_p(T)H$ and magnetization (*AC* and *DC*) were carried out using Quantum Design *PPMS* (Physical Property Measurement System).

## Results and Discussion

Fig. 1 depicts the room temperature X-ray diffraction (*XRD*) plot of the studied $Bi_3Ni$. The resultant compound is mainly single phase, with minor impurity of Bi at 27.2$^0$. The fitted and observed Reitveld parameters are shown in Table 1. $Bi_3Ni$ crystallizes in orthorhombic structure with space group *Pnma*, The lattice parameters are $a$ = 8.878(5)Å, $b$ = 4.102(1)(5)Å, and $c$ = 11.479(1)Å. These lattice parameters are in agreement with earlier reports [10-14]. For $Bi_3Ni$, there are three Bi sites namely $Bi_1$, $Bi_2$ and $Bi_3$ having fixed y at ¼ and varying x and z coordinates [13]. For Ni as well the y is fixed at ¼ and x and z coordinates are varied. So in effect there are four coordinate sites i.e., three for Bi and one for Ni in unit cell of $Bi_3Ni$. Further, one unit cell of $Bi_3Ni$ consists of four sub unit cells. The representative unit cell being determined from Reitveld analysis of the studied $Bi_3Ni$ is given in Fig. 1(b).

The *DC* and *AC* magnetic susceptibility plots of studied $Bi_3Ni$ are shown in Fig. 2 and 3 respectively. Namely, Fig. 2(a) depicts the DC magnetic susceptibility ($\chi$) in both zero-field-cooled (*FC*) and field-cooled (*FC*) situations in temperature range of 2K to 10K. The applied field is 10Oe. Superconductivity is observed at 4.1K with a sharp diamagnetic transition in magnetic susceptibility ($\chi$) in both *ZFC* and *FC* situations. The superconducting volume fraction seems to be around 87.6% as calculated from *FC* ($\chi$). This is slightly higher than as reported in ref. 13.Though an estimated value is given, still we believe estimating superconducting volume fraction without exactly knowing the pinning properties is not correct. What one can safely conclude from Fig. 2(a) is that the studied $Bi_3Ni$ is a bulk superconductor with superconducting transition temperature ($T_c$) at 4.1K. The *AC* susceptibility of $Bi_3Ni$ in both real ($\chi^{/}$) and imaginary ($\chi^{//}$) parts at frequency of 333Hz and amplitude of 1.0Oe is shown in Fig. 2(b). We also studied the AC susceptibility of $Bi_3Ni$ in both real ($\chi^{/}$) and imaginary ($\chi^{//}$) parts at various amplitude of 3-11 Oe and fixed frequency of 33 Hz. These results are depicted in Fig. 3. Basically there is nearly no shift in the peak position temperature with increase in *AC* drive field. This is unlike high $T_c$ superconductors (*HTSc*) [17] or the recently invented pnictides [18]. Fundamentally this shows that the superconducting grains are strongly coupled in $Bi_3Ni$ and hence inter/intra grains superconductivity characteristic of this compound must be good enough for practical applications.

The isothermal magnetization $M(H)$ plots for studied $Bi_3Ni$ at various fields and temperatures are depicted in Figs. 4(a-d). The low field (< 200Oe) isothermal $M(H)$ plots of $Bi_3Ni$ at $T$ = 2, 2.5, 3 and 3.5K are shown in Fig. 4(a). The inversion of these $M(H)$ marks the lower critical field ($H_{c1}$) for studied $Bi_3Ni$ superconductor. The $H_{c1}$ is around 150Oe at 2K and decreases monotonically with increase in $T$ to 30Oe at 3.5K. These values of $H_{c1}$ for studied $Bi_3Ni$ are in good agreement with an earlier report [13, 14].

Fig. 4(b) depicts the isothermal $M$ ($H$) plot for real part of *AC* susceptibility ($M^{/}$) with applied field of up to 5 kOe at 2K. This is done to know the extent of complete flux inclusion into the superconductor. This field roughly coincides with the upper critical field ($H_{c2}$) of the superconductor at the particular temperature. As seen from Fig. 4(b), the complete inclusion of the flux takes place at above 3 kOe. The 10%, 50% and 90% of the same is marked in the Fig. 4(b). By definition one takes the upper critical field ($H_{c2}$), to be 50% of externally excluded field, instead of full exclusion i.e. 100%. This is done in accordance with ref. 13. The upper critical field value of $Bi_3Ni$ thin films [13] is slightly higher than our value for studied bulk $Bi_3Ni$ superconductor.

Complete (all four quadrants) isothermal $M(H)$ loops for studied $Bi_3Ni$ are shown in Fig. 4(c) at T = 2, 2.5, 3 and 3.5K with in +3000 to – 3000Oe applied fields. The $M(H)$ plots exhibit expected symmetric superconducting loops with their closing at above 2000 Oe. This means the irreversibility field ($H_{irr}$) is close to 2000Oe. Interestingly a careful close look of these plots shows some magneto-superconductivity like asymmetry. To elaborate on this, we zoom Fig. 5(c) and the result is shown in Fig. 4(d). It is clear from Fig. 5(d) that all the superconducting *M*(*H*) loops are closed below and above base line in first and fourth quadrants respectively. This is unusual, because in normal course (with out magnetic background) the isothermal *M*(*H*) loops for a superconductor meet/close at the base/zero line. The external field matching with closing of the isothermal *M*(*H*) loops corresponds to the irreversibility field ($H_{irr}$), which is at around 2000Oe at 2K and is marked in Fig. 4(d). Hence as far as the irreversibility field ($H_{irr}$) is concerned, the same is around 2000Oe at 2K and is clear from both Fig. 4(d).

The moot question remains is that whether the slight asymmetry in *M*(*H*) loops is due to magnetic background or some other possible reasons like small grain size. To look for the possibility of either intrinsic (Ni in $Bi_3Ni$ structure magnetic) or extrinsic (un-reacted Ni) magnetism in studied $Bi_3Ni$, we carried out the moment versus temperature *M*(*T*) experiments and the results are shown in Fig. 5. The *M*(*T*) is carried out at an applied field of 100 Oe in both Zero-field-cooled (*ZFC*) and Field-cooled (*FC*) situations. The temperature range is 2-300K. As evidenced from the *M*(*T*) results of Fig. 5, studied $Bi_3Ni$ is superconducting below 4.1K with sufficient superconductivity shielding fraction and the normal state i.e. 4.1 to 300K, is non-magnetic. This is interesting, because despite being full occupancy of Ni in studied compound $Bi_3Ni$, no sign of magnetic ordering or even the moment of Ni is seen. The normal state *M*(*T*) of studied $Bi_3Ni$ clearly demonstrates that Ni is non magnetic in $Bi_3Ni$ structure. This is consistent with our *DFT* calculations.

To further investigate the normal state magnetism of Ni in $Bi_3Ni$, we did isothermal magnetization *M*(*H*) on studied $Bi_3Ni$. The *M*(*H*) done at 200K in -3000 to +3000 Oe in all four quadrants is depicted in inset of Fig. 5. Clearly the *M*(*H*) shown in inset of Fig. 5 is linear and thus rules out any magnetic ordering. Further the moment is also very small i.e., 0.002 emu/g at 3000Oe field. The calculated moment per Ni atom is in fact negligible ($10^{-4}$ $\mu_B$). At this point we conclude Ni is non-magnetic in $Bi_3Ni$ superconductor. This is similar to another Ni based superconductor $MgCNi_3$ [16]. As far as the small asymmetry of *M*(*H*) loops (Fig. 4d) is concerned, the same can be caused by grains size distribution as well. Remember, the presently studied sample of $Bi_3Ni$ is synthesized by an easy versatile method of vacuum encapsulation instead of the arc melting [10-12]. A representative *SEM* (scanning electron microscope) picture of the studied $Bi_3Ni$ compound is shown in Fig. 6. The typical grain size is laminar slab like and quite inhomogeneous, and this could be the reason behind slightly asymmetric $M(H)$ plots shown in Fig. 4(d). We recently observed a similar effect in particle size controlled $La_{1.8}Sr_{.15}CuO_4$ superconductor [19]. Because we did not find any conclusive evidence for magnetic contributions from Ni in $Bi_3Ni$ superconductor, hence we believe the slight asymmetry in $M(H)$ plots is due to inhomogeneous grain size and not due to magneto-superconductivity.

The temperature dependence of specific heat *C(T)* is measured in zero field and for 1 Tesla applied on Quantum design *PPMS* and is shown in Figure 7. The jump in electronic specific heat anomaly is clearly visible in zero magnetic field at around 4.2K. The value of jump is found to be $6.23\times10^{-2}$J/mol-K, see upper inset of Fig. 7. The normalized value of jump $(C_{es}-\gamma T_c)/\gamma T_c$ is found to be 1.50 that is close to *BCS* value of 1.43. The low temperature specific heat was fitted to Somerfield-Debye expression

$$C(T) = \gamma T + \beta T^3 + \delta T^5 \qquad \text{.......... (1)}$$

where $\delta T^5$ term represents the anharmonic contribution. From this fitting the values of Sommerfeld constant ($\gamma$) and $\beta$ are obtained. The $\gamma$ and $\beta$ give the value of electronic density of states and approximate value of Debye temperature respectively. The values obtained are $\gamma$ = 11.08mJ/mol-K$^2$, $\beta$ = 3.73mJ/mol-K$^4$ and $\delta$ = 0.0140*mJ*/mol-*K*$^6$. The fitting is shown in lower inset of Fig. 8. These values are in good agreement with other reported values [13]. The small difference in the values can possibly be because of little Bismuth impurity in our sample. From the value of $\beta$ we calculated the value of Debye temperature using $\Theta_D = (234zR/\beta)^{1/3}$ here $z$ being number of atoms per formula unit and $R$ is gas constant. The value of $\Theta_D$ is found to be 127.77K. From the value of Sommerfeld constant we have calculated value of electronic Density of states at Fermi level $N(E_f)$ using formula:

$$N(E_F) = \frac{3\gamma}{\pi^2 K_B^2} \qquad \text{.......... (2)}$$

The value of is $N(E_f)$ found to be 4.697 states/eV-*fu*.

To further investigate the magnetic nature and detailed microscopy of this compound we perform the density functional based calculations using the Full Potential Linear Augmented Plane Wave (*FP-LAPW*) as implemented in *WIEN2k*. To describe the exchange and correlation in the crystal Hamiltonian, we use the local density approximation (LDA). The *LAPW* sphere radii for Bi and Ni were chosen as 2.50 a.u. and 2.00 a.u., respectively. The total energy was converged for 96 k-points in the irreducible *BZ*. The experimental lattice parameters were used. The unit cell (u.c.) consists of four formula units (*fu*) with total of 16 atoms (4-Ni and 12-Bi). Each Ni atom is surrounded by six Bi atoms and two Ni atoms. We relaxed the internal coordinates of the Bi ions via force minimization technique to a maximum force of 1mRy/a.u. Force minimization to determine the lattice coordinates yields consistent values with the experiments. In Fig.8(a), we show the electronic density of states (*DOS*) of $Bi_3Ni$ for the fully relaxed structure. As revealed in Fig.8(a), we find that the Ni-*d* and Bi-*p* states are distributed over the large energy range suggesting strong covalent bonding in Ni-*d* and Bi-*p* states. The bands in lower valance band (-13.8 to -9.6eV) consists mainly of Bi-*s* states with small contribution of Ni states. Then follows a gap in *DOS* of about 3.6eV and upper part of valance band (-6.0eV to Fermi level) is dominated by Ni-*d* states. The uniform spread of *DOS* over large energy range suggests covalent nature of bonding. To further investigate the details of Ni states, we have calculated the *l*-resolved Ni projected density of states (*PDOS*), which shows that most of Ni states come from Ni-*d* orbitals. At Fermi level the total states are dominated mostly from Ni-*d* with relatively smaller contribution from Bi-*p* states. The value of *DOS* at Fermi level is around 10.2 States/eV-u.c. (2.55 states/eV-*fu*), which shows that system is metallic. Non magnetic nature is confirmed from total *DOS* for spin up and spin down states bring overlapping exactly. Since spin-orbit coupling (*SOC*) plays important role in Bi based compounds, we have done the same calculations by incorporating *SOC*. The stabilization of the energy with *SOC* was lower by ~35.47*mRy* per atom, which is quite large. However this is expected as bismuth has strong *SOC* [20]. The Fermi level is on a broad peak (see inset of Fig. 8a), ruling out any possibility of doping dependent enhancement in $T_c$ as per *BCS* criteria. The value of electronic density of states at Fermi level $N(E_F)$ with *SOC* is decreased to 2.11 States/eV/*f.u*. The main features of *DOS* remain unaltered by *SOC*.

From the value of calculated DOS at Fermi level and that obtained from low temperature specific heat we have estimated the value of electron phonon coupling constant λ using relation

$$\frac{\gamma_{expt}}{\gamma_{calc}} = 1 + \lambda \qquad \text{.........} (3)$$

From equation (2) we have

$$\gamma = \frac{\pi^2}{3} N(E_F) K_B^2 \qquad \text{.......... (4)}$$

Thus equation (3) is equivalent to

$$\frac{N_{expt}(E_F)}{N_{calc}(E_F)} = 1 + \lambda \qquad \text{.......... (5)}$$

Which gives $1 + \lambda = 1.842$, whence λ=0.842 that shows $Bi_3Ni$ is an intermediate coupling superconductor. When we use $N_{calc.}(E_F)$ being obtained considering *SOC*, the λ comes out to be ≈1.23. It is known that *BCS* superconductors can be well described using MacMillan Formula:

$$T_C = \frac{\Theta_D}{1.45} exp\left[-\frac{1.04(1+\lambda)}{\lambda-\mu^*-0.62\lambda\mu^*}\right] \qquad \text{............ (6)}$$

where $\Theta_D$ is Debye temperature and is determined from specific heat measurements and μ* is Coulomb coupling constant. The standard value of μ*=0.13 gives a $T_c$ of 4.5 K for λ=0.842 which is in good agreement with corresponding experimental value of 4.1K. However for the case with *SOC*, λ=1.23 (as should be actually considered), the $T_c$ is overestimated to ~8.6K for μ*=0.13, which is more than double of the experimental value. To obtain experimental $T_c$ one has to take μ*~0.26, which is quite a large value. It is interesting to see the larger value of μ* to reproduce experimental $T_c$ in case of *SOC*. Such a large value for μ* have been recently reported for solid picene ($C_{22}H_{14}$) [22] by Subedi et. al. Thus dependence of μ* on *SOC* is interesting to investigate in future.

The non magnetic character of $Bi_3Ni$ is also evident from Stoner condition i.e., $I*N(E_F)$, I is Stoner parameter [21], $N(E_F)$ is total density of states at Fermi level due to *d* states of magnetic elements, which here are 4-Ni atoms. We found $I*N(E_F)$ here to be~0.55 per unit cell, which for per Ni atom is ~0.136. For ferromagnetism the value must be close to or greater than 1, which is not the case for $Bi_3Ni$. The value is so small (comparable to Al with 0.126 [21]) that any possibility of ferromagnetism in this compound can be overruled. To further ensure the possibility of any local minima near zero magnetic moment, we did fixed spin moment (*FSM*) calculations by fixing total magnetic moment on the unit cell and observe corresponding energy. The so obtained magnetic energy is then fitted to sixth order Ginzburg Landau (GL) equation $E = \frac{1}{2}(Am^2) + \frac{1}{4}(Bm^4) + \frac{1}{6}(Cm^6)$ that gives us *A*=1.18±0.022mRy/$\mu_B^2$, *B*=0.155±0.0.035mRy/$\mu_B^4$, *C*=-0.02053±0.0091 *mRy*/$\mu_B^6$ per Ni atom. Thus fitted plot for moment per unit cell (4 *fu* i.e. 4Ni) versus energy is shown in Fig. 8(b). These values suggest that system does not exhibit any signatures of ferromagnetism.

Summarily, we synthesized $Bi_3Ni$ superconductor by an easy solid state synthesis route with $T_c$ of 4.1K. Various physical properties of the compound including structure, *AC*/*DC* magnetization and heat capacity etc. are presented and discussed. Detailed specific heat analysis and *DFT* calculations revealed that this compound is having moderate electron-phonon coupling constant of around 1.23 and Ni is non magnetic. Further it is concluded that ferromagnetic interactions play no role in superconductivity of $Bi_3Ni$. Further, the value of $\mu^*=0.26$ is required to reproduce experimental value of $T_c$ as per McMillan equation, while considering the strong *SOC* character of Bi. This adds to the future scope to investigate the detailed *SOC* dependence of $\mu^*$ in $Bi_3Ni$.

## Acknowledgement

Jagdish Kumar, Anuj Kumar, and Arpita Vajpayee would like to thank the *CSIR* for the award of Senior Research Fellowship to pursue their *Ph. D* degree. Authors thank Prof. R.C. Budhani, DNPL and Dr. Hari Kishan, HOD for their keen interest and encouragement for superconductivity research.

Table 1 Rietveld refined structure parameters of $Bi_3Ni$

[$a$ = 8.878(5)Å, $b$ = 4.102(1)(5)Å, $c$ = 11.479(1))Å, $R_p$ = 7.03, $R_{wp}$= 9.88, $R_{exp}$= 3.50,$\chi^2$= 2.37 , Vol = 418.081$A^3$]

| Atom | x | y | z |
|---|---|---|---|
| Bi1 | 0.293(5) | 1/4 | 0.890(8) |
| Bi2 | 0.378(2) | 1/4 | 0.588(7) |
| Bi3 | 0.407(1) | 1/4 | 0.177(6) |
| Ni | 0.069(5) | 1/4 | 0.513(2) |

## Figure Captions

Figure 1(a) X-ray diffraction (XRD) pattern of studied $Bi_3Ni$, the impurity Bi line is marked.

Figure 1(b) Schematic unit cell of $Bi_3Ni$.

Figure 2(a) *DC* magnetic susceptibility $M(T)$ in *ZFC* (Zero-Field-Cooled) and *FC* (Field-Cooled) situations at 10Oe for $Bi_3Ni$.

Figure 2(b) *AC* magnetic susceptibility in both real ($M^{/}$) and imaginary ($M^{//}$) situations at 333Hz and 1Oe amplitude for $Bi_3Ni$.

Figure 3 *AC* magnetic susceptibility in both real ($M^{/}$) and imaginary ($M^{//}$) situations at fixed frequency of 33Hz and varying Amplitudes of 3-11Oe for $Bi_3Ni$.

Figure 4(a) Isothermal magnetization (*MH*) plots at 2, 2.5, 3 and 3.5K in low field range of <200Oe for $Bi_3Ni$, the lower critical field ($H_{c1}$) is marked.

Figure 4(b) Isothermal magnetization (*MH*) for real part of *AC* susceptibility ($M^{/}$) with applied field of up to 5 kOe at 2K $Bi_3Ni$, the upper critical field ($H_{c2}$) is marked.

Figure 4(c) Isothermal magnetization (*MH*) plots at 2K in high field range of up to 2500Oe in four quadrants for $Bi_3Ni$.

Figure 4(d) Expanded *MH* plots at 2K in high field range of up to 2500Oe in four quadrants for $Bi_3Ni$, the $H_{irr}$ is marked.

Figure 5 $M(T)$ in both *ZFC* and *FC* situations at 10Oe for $Bi_3Ni$ in temperature range of 2 to 300K, the inset shows the $M(H)$ at 20K with in 3000Oe in all four quadrants.

Figure 6 Typical Scanning Electron Microscope (*SEM*) picture of studied $Bi_3Ni$.

Figure 7 Specific heat versus temperature plot $C_P(T)$ in temperature range of 2-100K for studied $Bi_3Ni$, the upper and lower insets show the electronic specific heat anomaly at $T_c$ and he fitting parameters respectively.

Figure 8(a) Calculated electronic density of states (*DOS*) for the fully relaxed structure $Bi_3Ni$, inset shows the *DOS* with and without *SOC* around Fermi level.

Figure 8(b) Fixed spin moment calculations of $Bi_3Ni$

Figure 1(a)

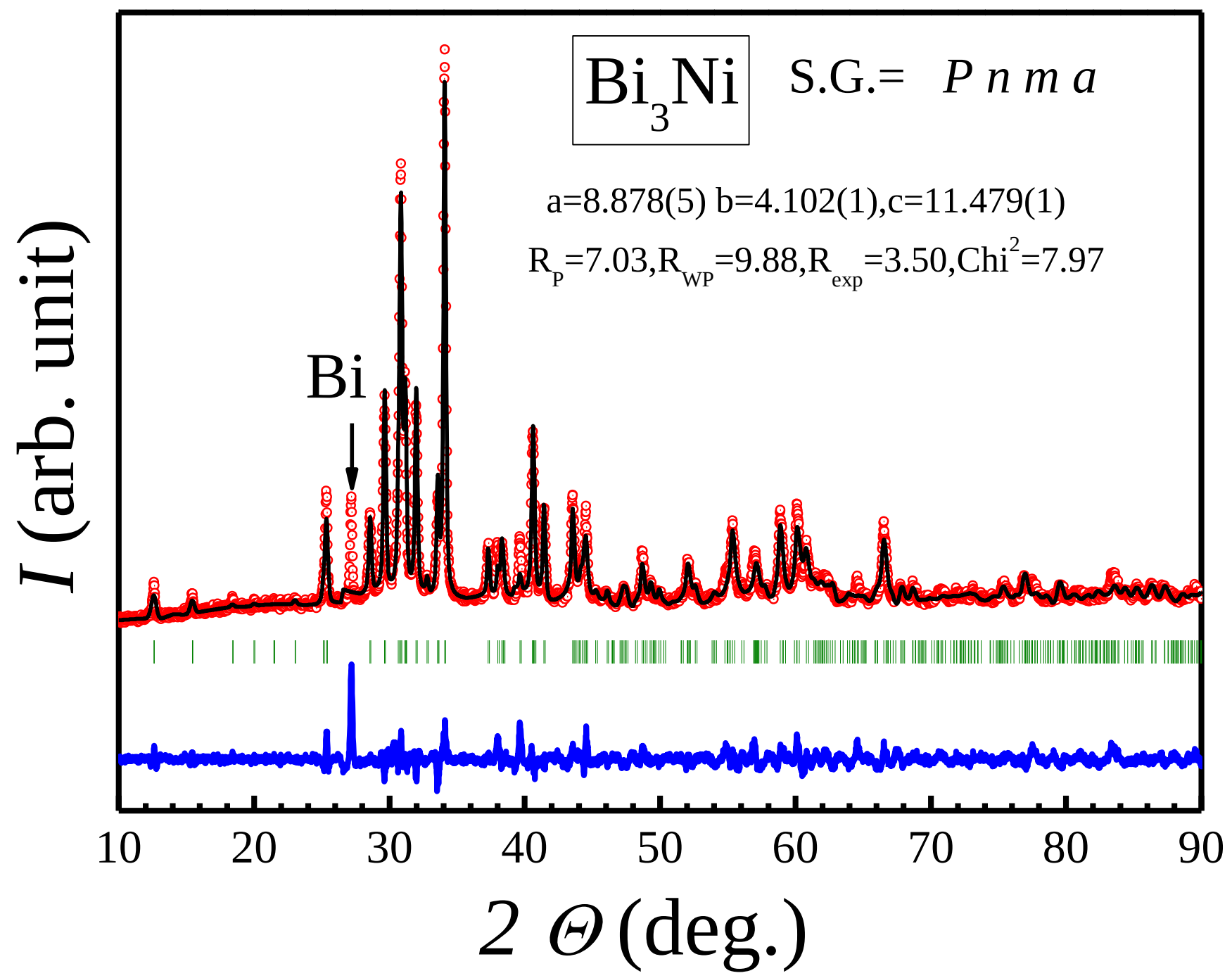

Figure 1(b)

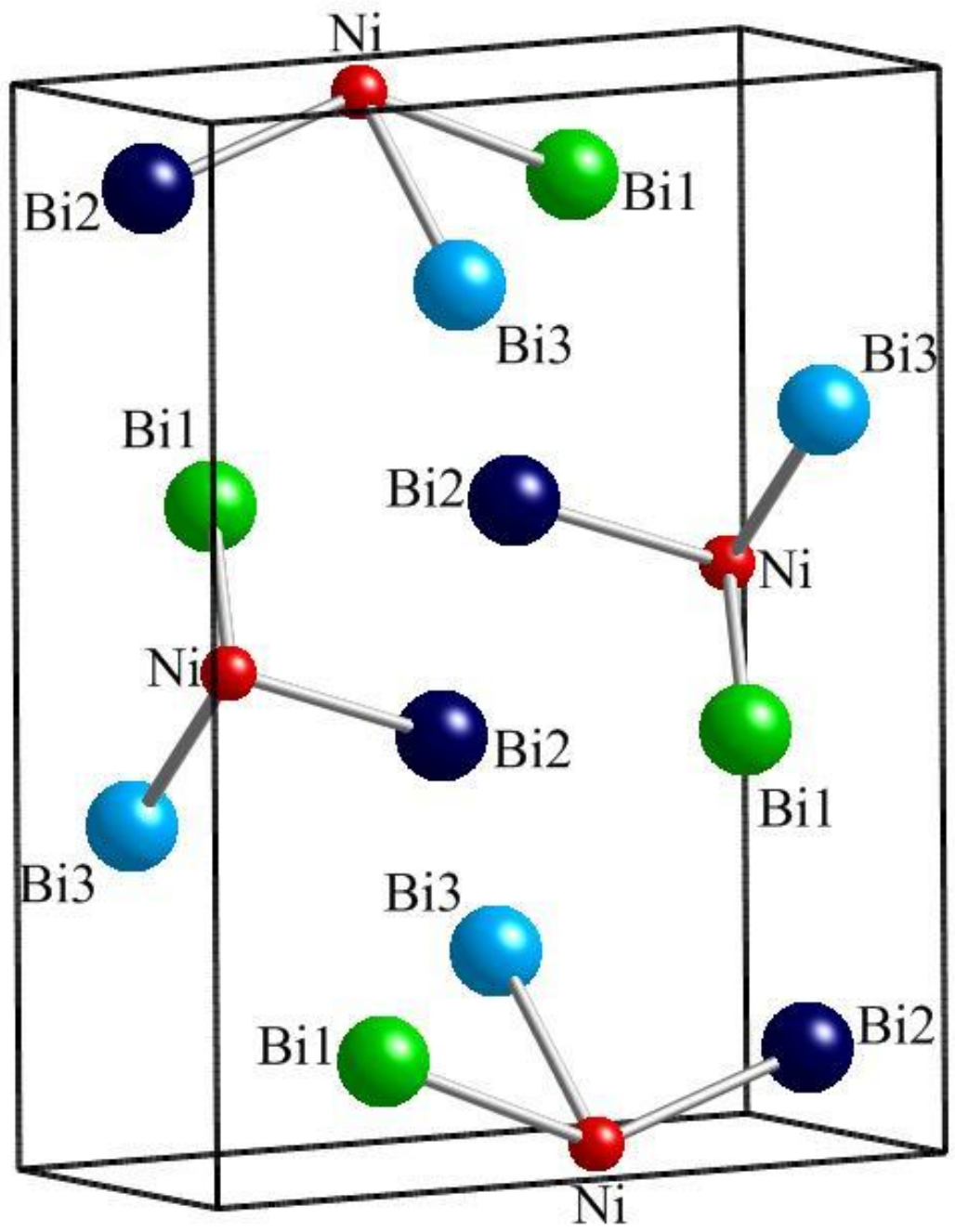

Figure 3 (a)

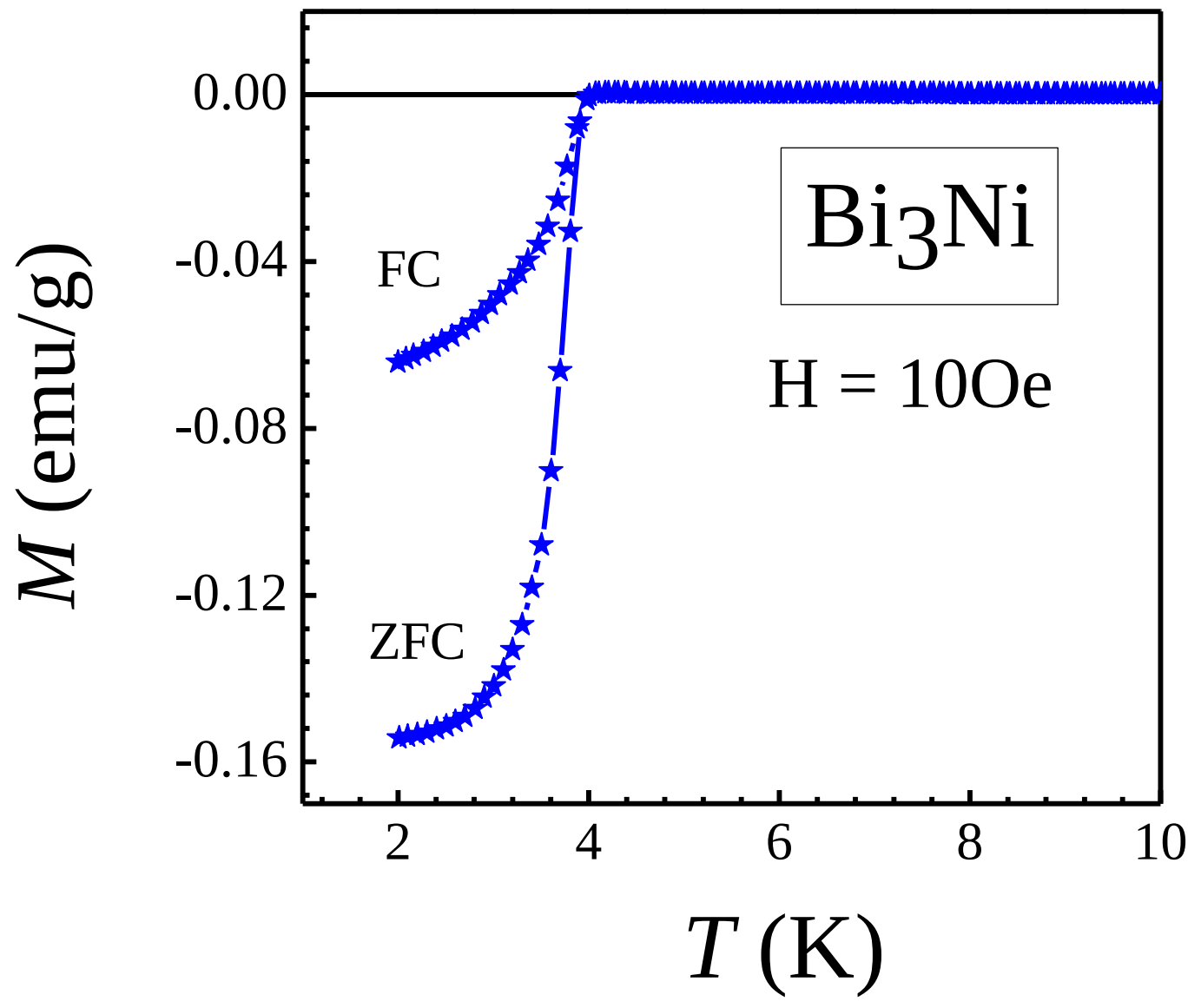

Figure 3 (b)

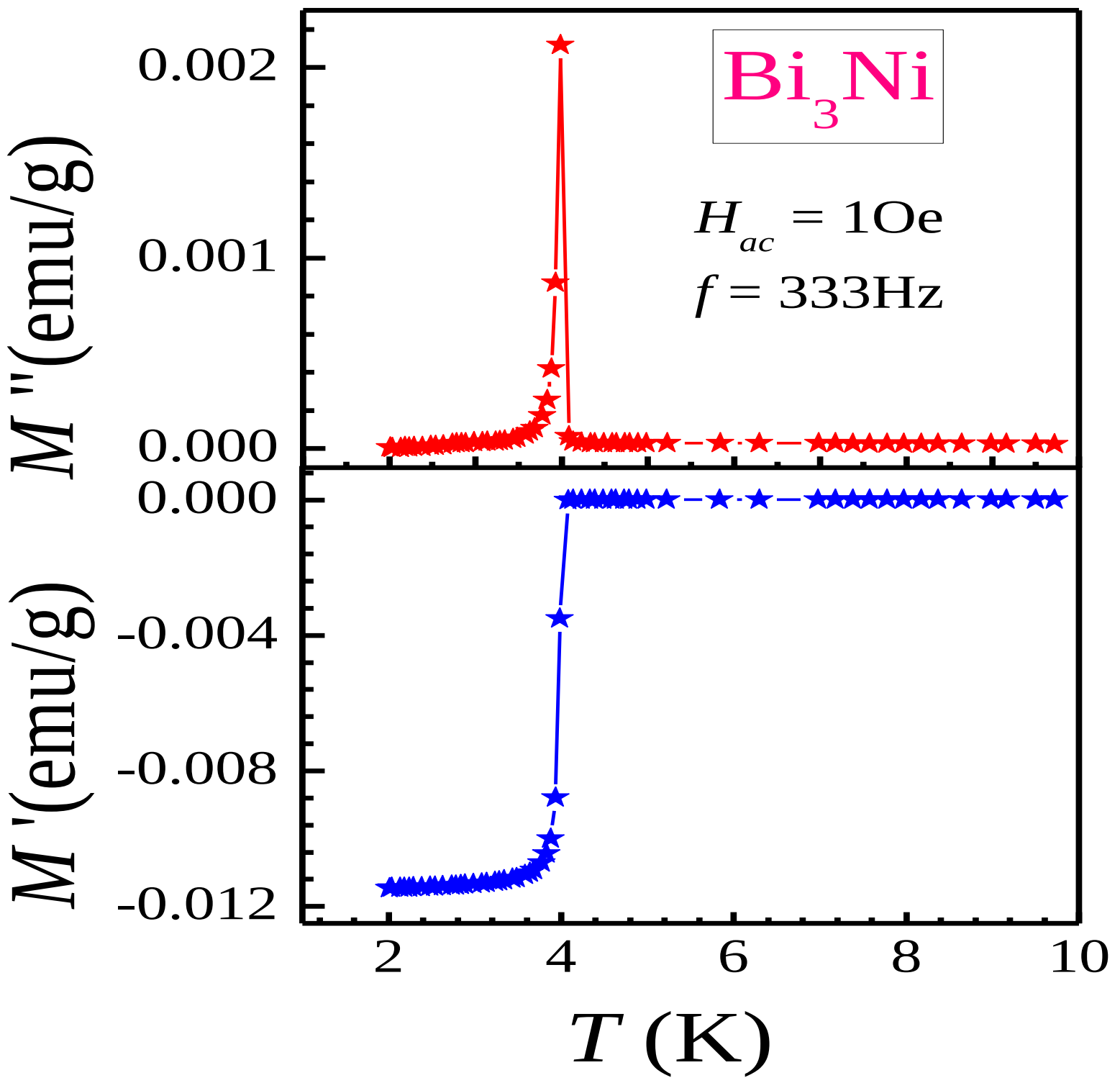

Figure 4

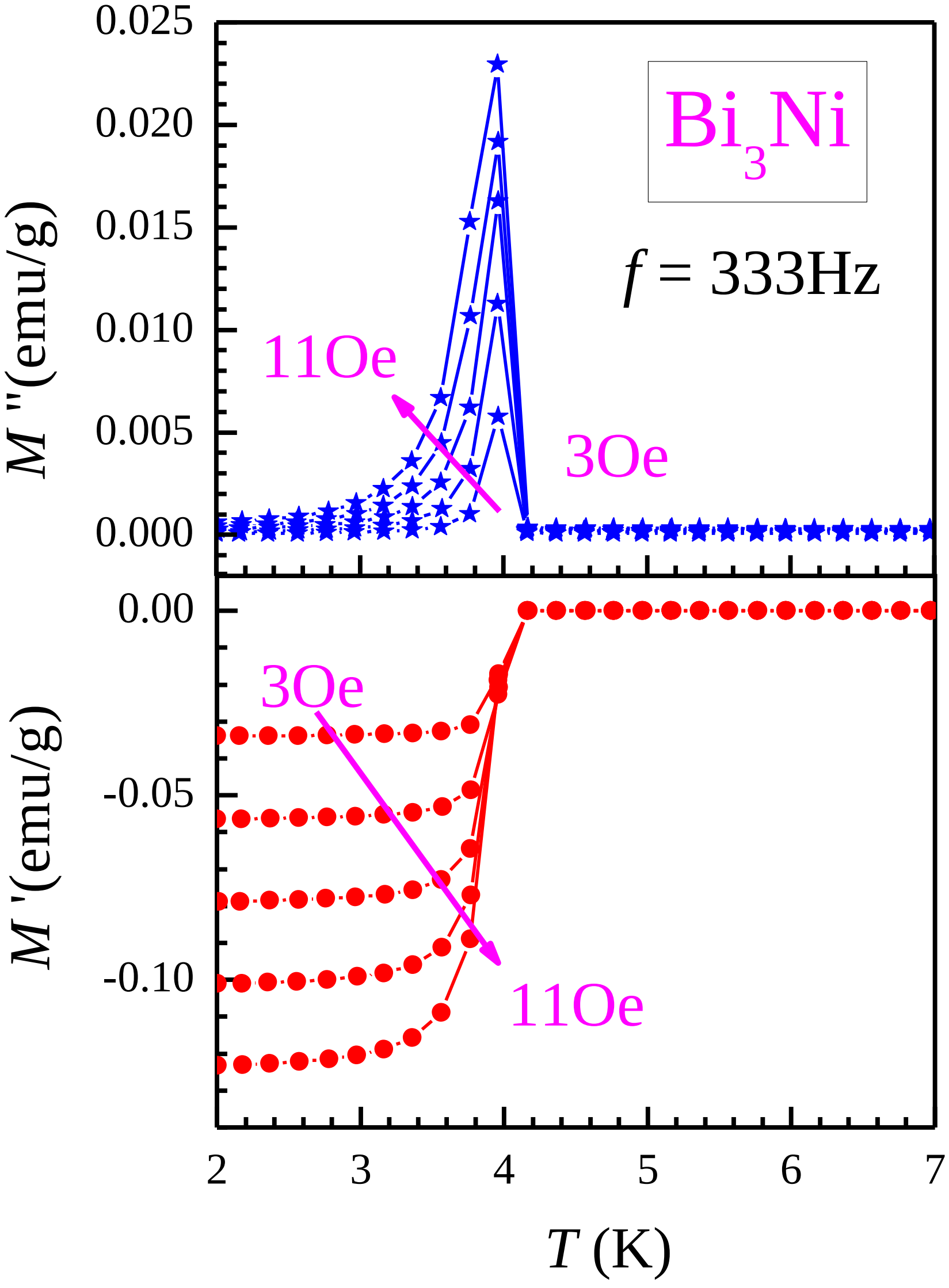

Figure 5(a)

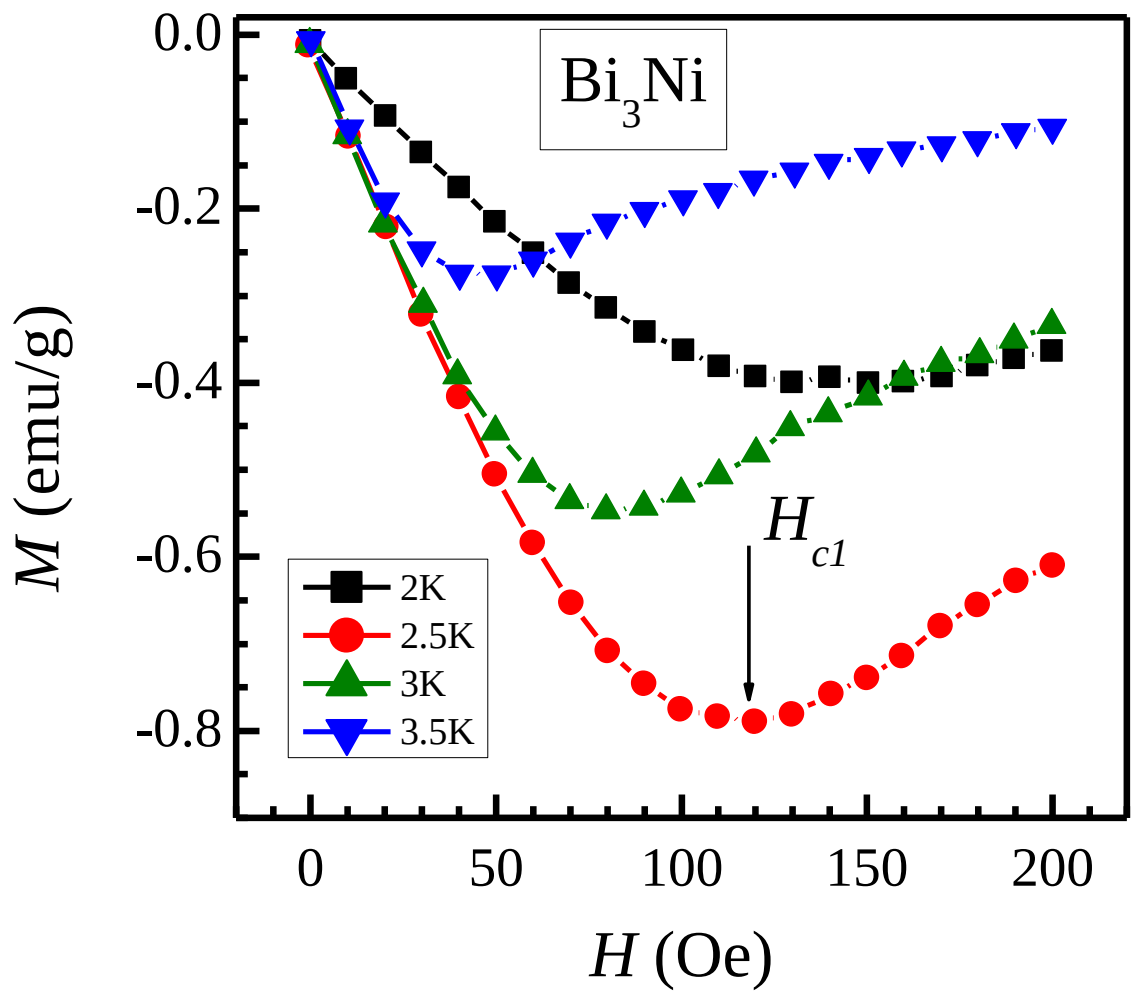

Figure 5(b)

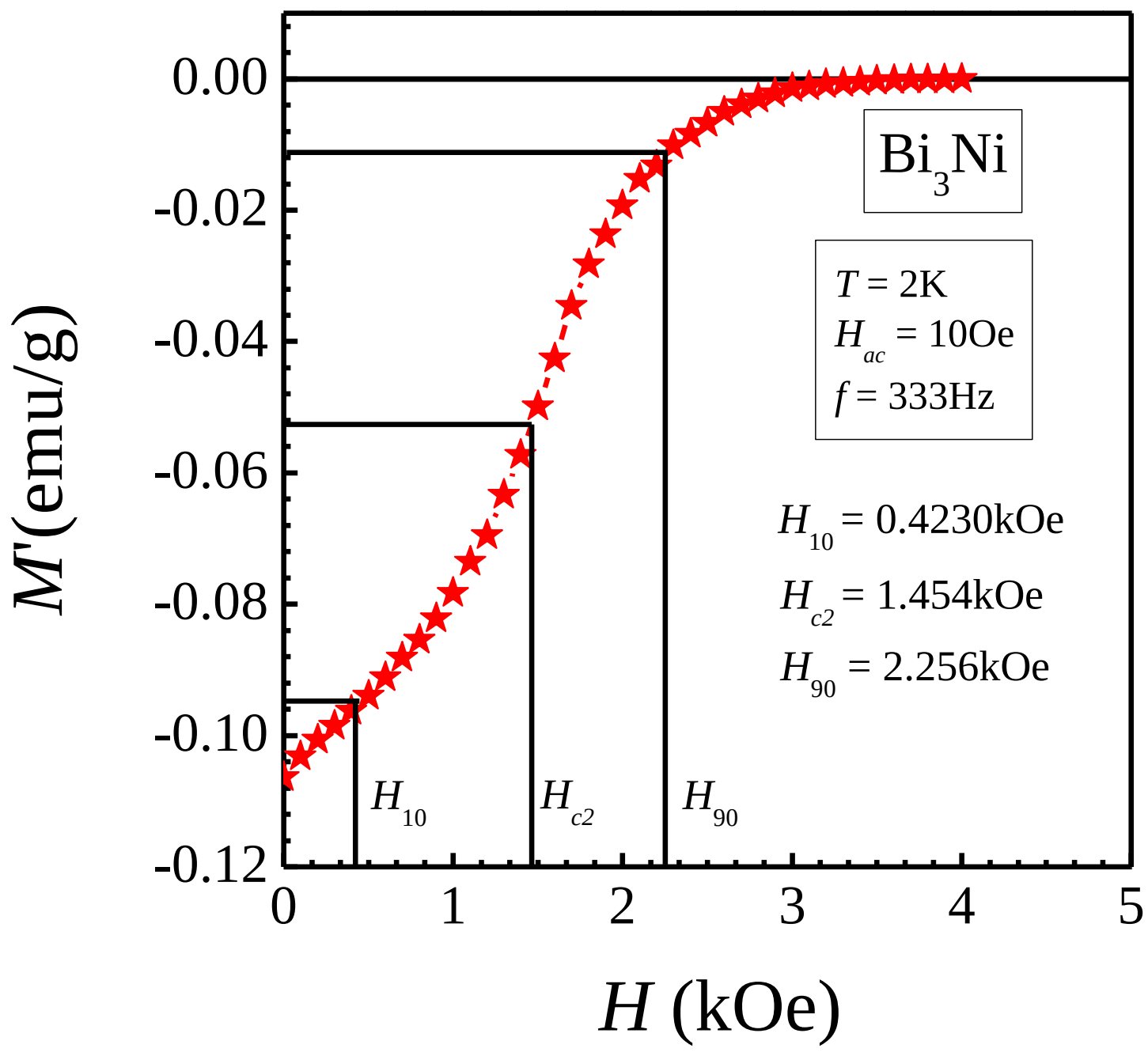

Figure 5 (c)

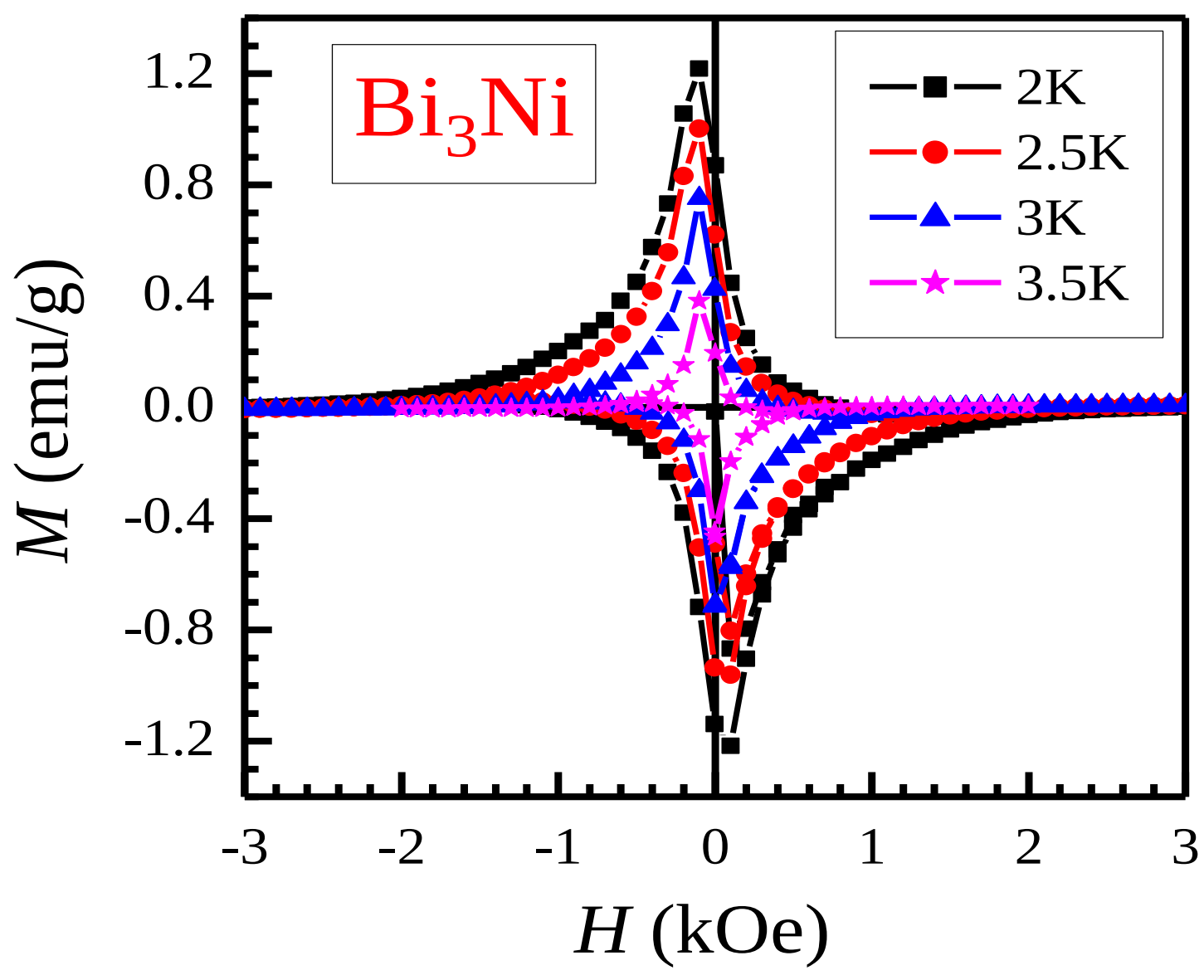

Figure 5(d)

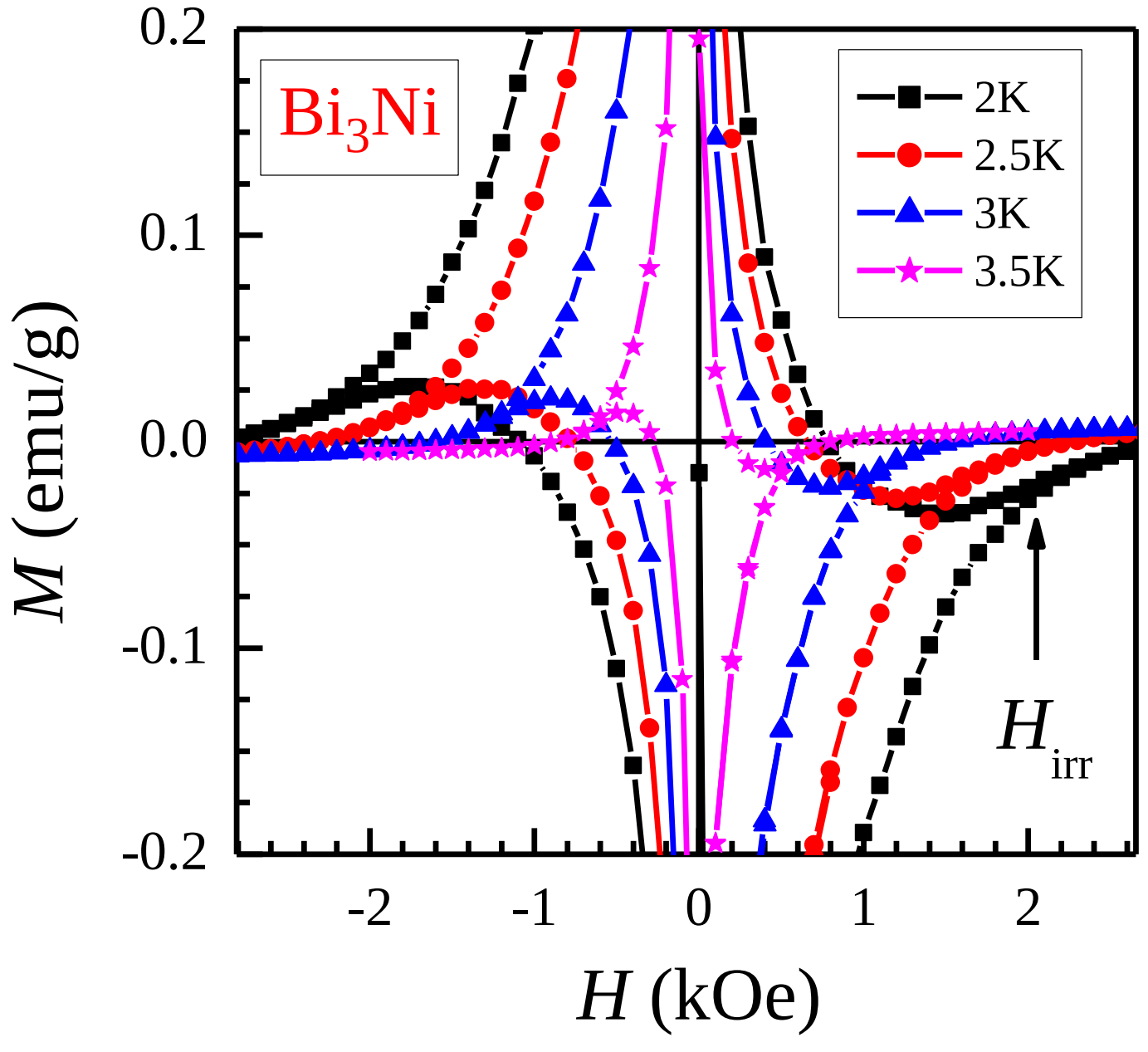

Figure 6

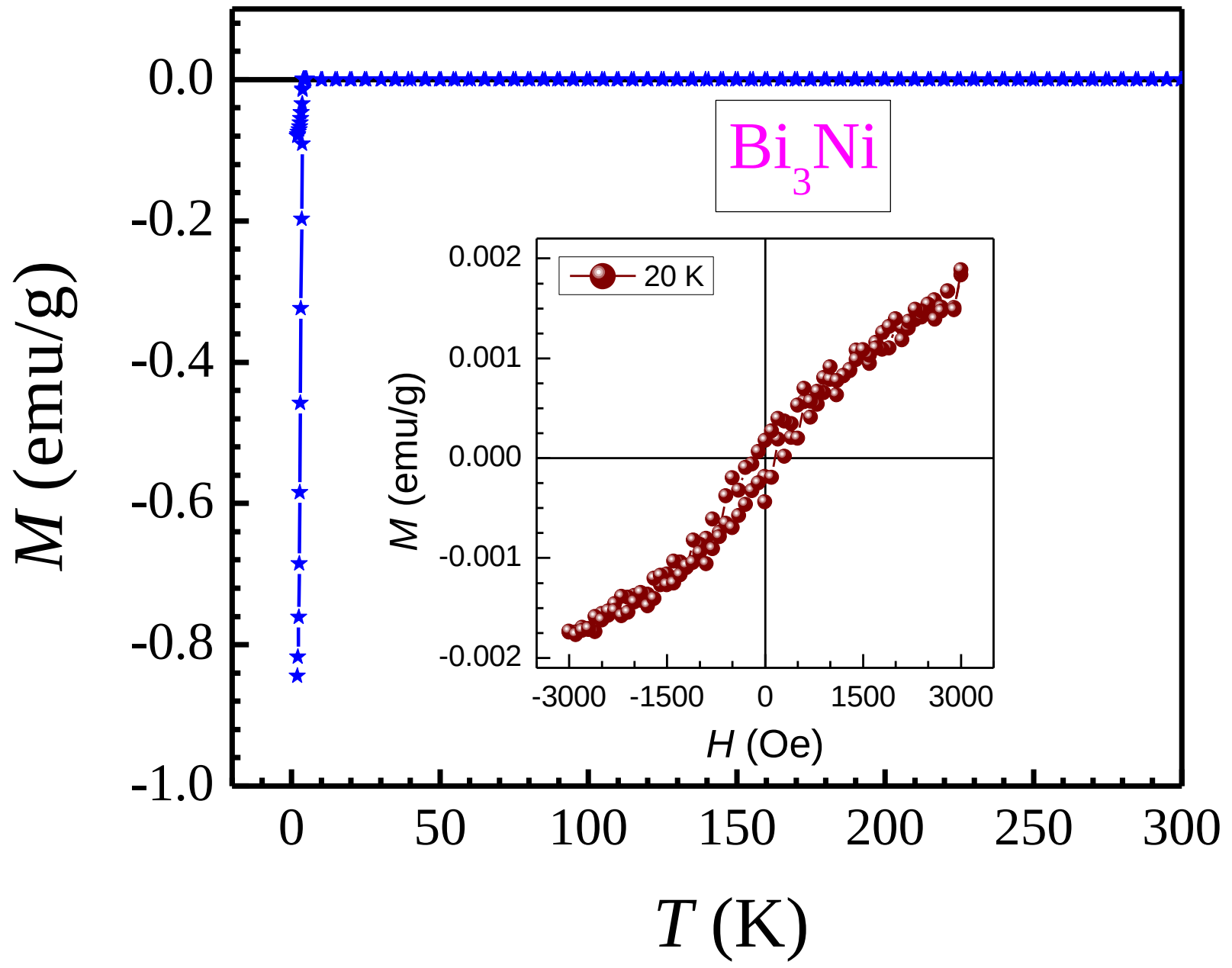

Figure 7

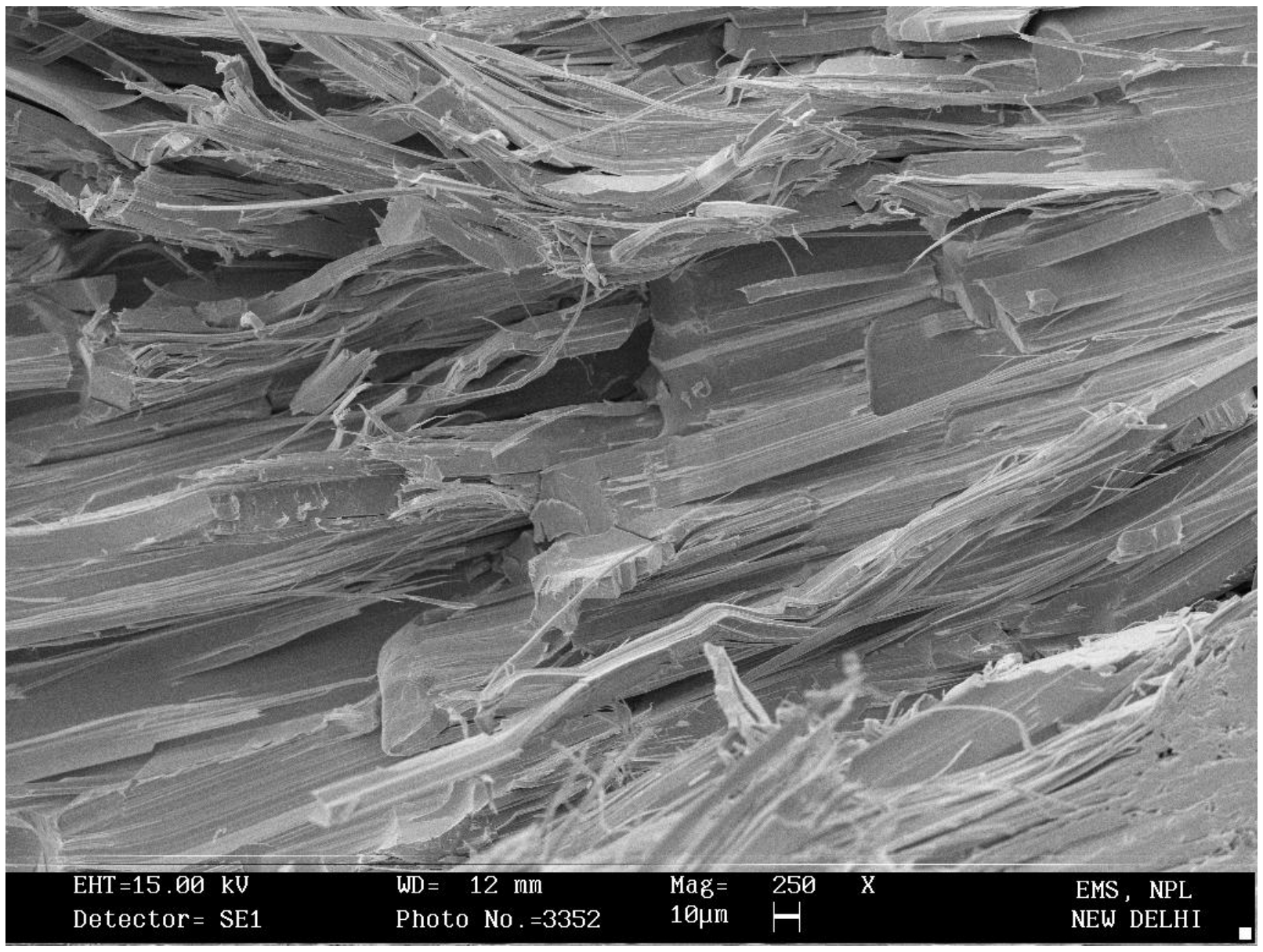

Figure 8

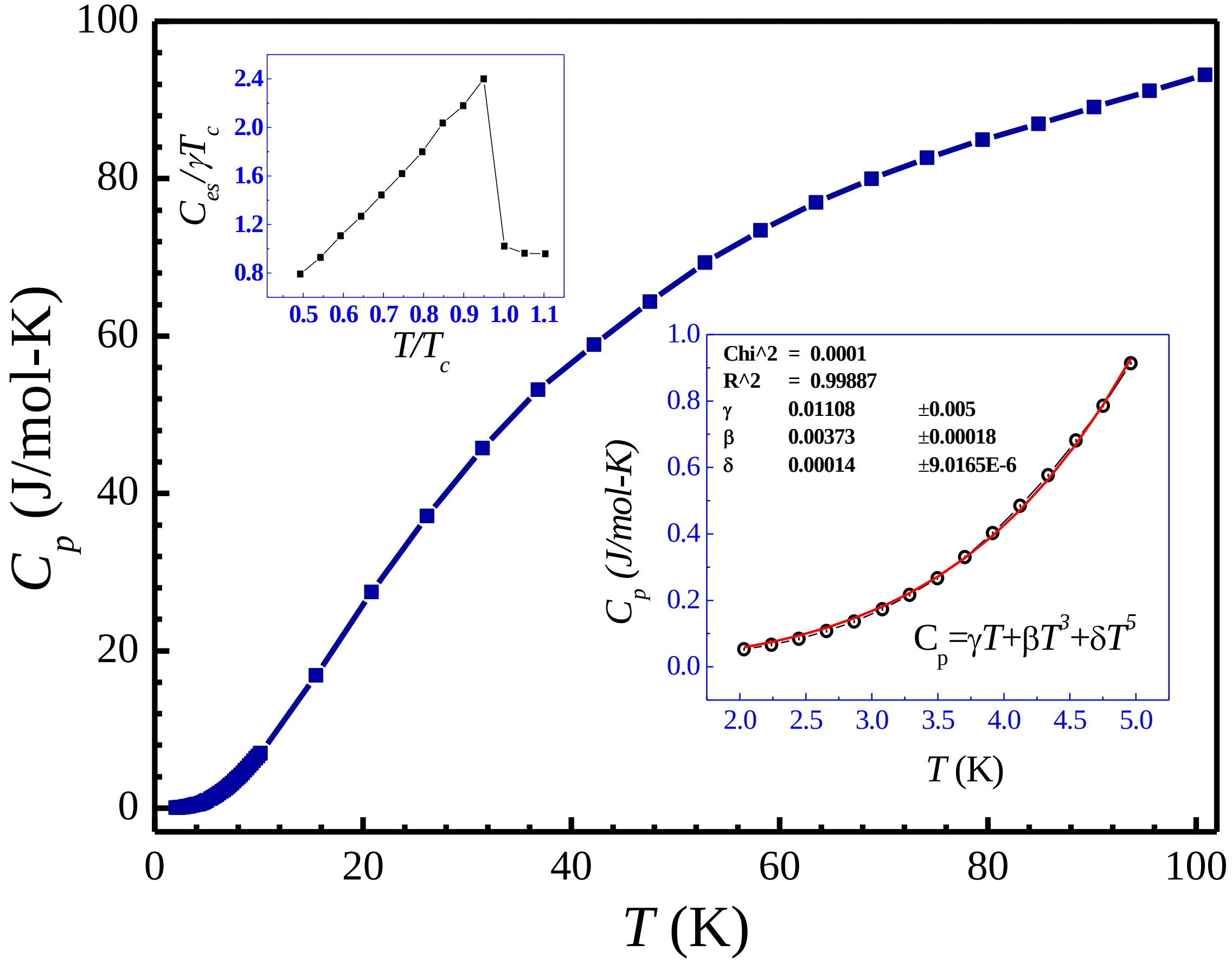

Figure 8(a)

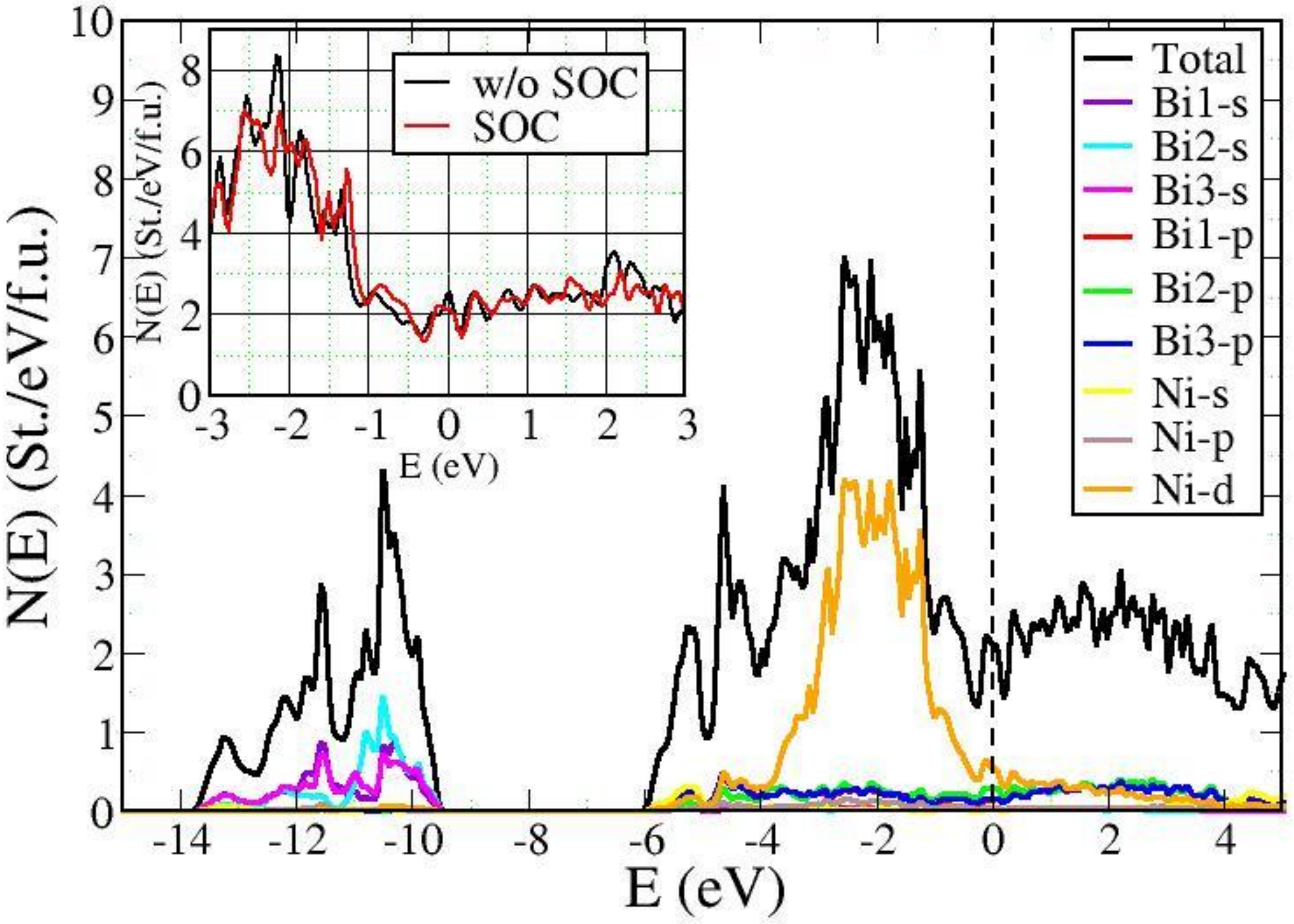

Figure 8(b)

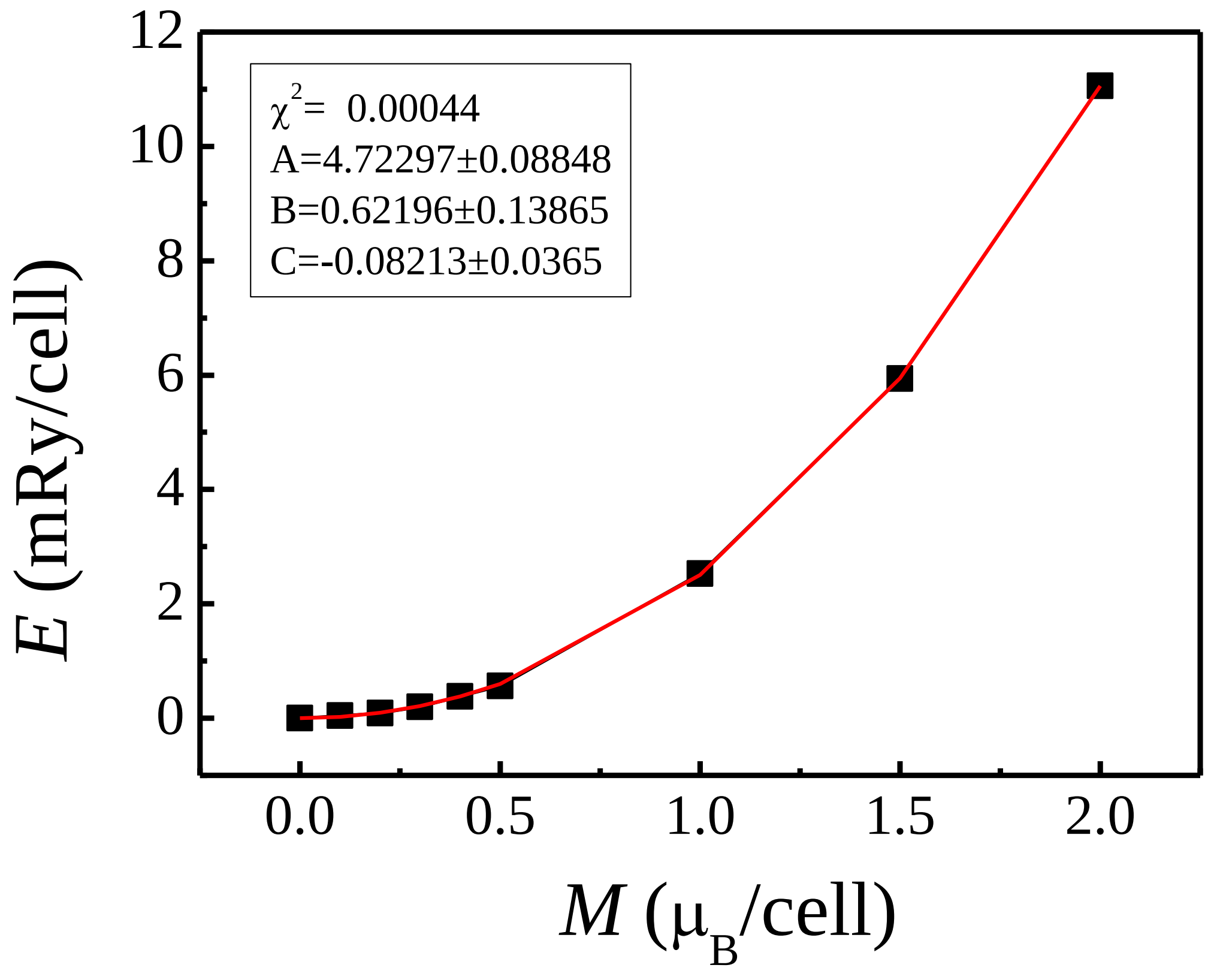